# Data Processing Protocol for Regression of Geothermal Times Series with Uneven Intervals

*Palash Panja[†1,2], Pranay Asai[2], Raul Velasco[1], Milind Deo[2]*
[†]ppanja@egi.utah.edu
[†]Corresponding author
[1] Energy & Geoscience Institute, University of Utah
[2] Department of Chemical Engineering, University of Utah,

**Abstract**

Regression of data generated in simulations or experiments has important implications in sensitivity studies, uncertainty analysis, and prediction accuracy. Depending on the nature of the physical model, data points may not be evenly distributed. It is not often practical to choose all points for regression of a model because it doesn't always guarantee a better fit. Fitness of the model is highly dependent on the number of data points and the distribution of the data along the curve. In this study, the effect of the number of points selected for regression is investigated and various schemes aimed to process regression data points are explored. Time series data i.e., output varying with time, is our prime interest mainly the temperature profile from enhanced geothermal system. The objective of the research is to find a better scheme for choosing a fraction of data points from the entire set to find a better fitness of the model without losing any features or trends in the data. A workflow is provided to summarize the entire protocol of data preprocessing, regression of mathematical model using training data, model testing, and error analysis. Six different schemes are developed to process data by setting criteria such as equal spacing along axes (X and Y), equal distance between two consecutive points on the curve, constraint in the angle of curvature, etc. As an example for the application of the proposed schemes, 1 to 20% of the data generated from the temperature change of a typical geothermal system is chosen from a total of 9939 points. It is shown that the number of data points, to a degree, has negligible effect on the fitted model depending on the scheme. The proposed data processing schemes are ranked in terms of $R^2$ and NRMSE values.

**Introduction**

Data collection, processing and interpretation have become pivotal tool to help making informed and risk evaluated decision for in every industry. The data collected via appropriate design of experiments and processed through various machine learning algorithms allow us to evaluate or optimize performance of any given setup/system. Previously, to evaluate or predict the behavior of any given system numerous physical/chemical experiments were carried out and the data was manually collected and analyzed. These methods were time and resource consuming and heavily relied on human accuracy. But today, most of the required data can be generated using highly tuned simulations combined with various machine learning algorithms which are custom built to include all the desired physics and chemistry or any other type of laws/interactions.

Stewart Robinson (Robinson 2004) explains the what why and when to use the simulations in his book and how it helps us in saving tremendous amount of time, energy and also the required amount of material for a given study. Simulation is a systematic tool, which when used correctly can be powerful in helping solve complex problems and aids in developing an optimized system. Machine learning techniques take us one step further by helping us develop complex mathematical models, by taking simulation data into account. Beylkin et al.(Beylkin, Garcke et al. 2009) developed an algorithms based on multivariate linear regression to develop modes for scattered data. These virtual mathematical models can be further optimized and developed to imitate the real models existing in the real world, hence eradicating the need of performing the cumbersome experiments. Apart from being robust, these models allow us to reduce the error (by defining the tolerance) and thus making them even more accurate. A list of time series forecasting models is provided later.

Usually while performing curve fitting in linear models the $R^2$ value is considered as the benchmark to establish the fitness of the curve, with $R^2$ approaching 1 being the best. However, this is not applicable for nonlinear curve fitting. Spiess and Neumeyer (Spiess and Neumeyer 2010) shows how $R^2$ is an inadequate measure to validate the fitness of curve in nonlinear models.

Simulations can be carried out to analyze the performance of a system over period of time and also to predict future behavior. This is especially in the field of oil and gas engineering and heat transfer in geothermal reservoir where simulation proves to be a most reliable tool. Studies performed by Okouma et al. (Okouma Mangha, Ilk et al. 2012), Kamari et al. (Kamari, Mohammadi et al. 2017) and Alom et al. (Alom, Tamim et al. 2017) on decline curve analysis for shale oil and shale gas, uses the similar technique to forecast the oil/gas production. The three major steps of any simulation based study involves design of experiment, simulation design and data interpretation. When the objectives of the study are established the first step is to design the experiment by choosing the number of parameters to be studied, their range and the different combinations to evaluate the performance of any individual parameters. This is usually done to make the model robust and efficient. Shaibu et. al. (Shaibu, Cho et al. 2009) highlights the importance of robust design, especially for a time oriented data to improve the accuracy of the model. The combinations are usually done by using various design of experiments (DOE) techniques like the Box Behnken method (Box and Behnken 1960). This kind of DOE helps in singling out the effect of individual parameters and helps us visualize the combined effect.

After the DOE is developed, the next step in simulation is the simulation design. In this step, as per the nature of study to be performed, a relevant simulator is chosen or can be self-designed (provided that it incorporates all the correct equations required for the study). Once the simulator is chosen, the setup is designed to represent the system in a virtual environment in the exact manner or by making certain reasonable assumptions. All the parameters to be studied are defined as per the simulator requirements and depending on the nature of study a time period is chosen (if it is a non-steady case). After setting all the things in the desired format (predefined by the simulator), the simulation is ran and the output data is collected. The last step in the simulation based study is to verify the accuracy of the output data. This can be done by performing few test experiments and using the same parameters to run the simulations. Then the simulation results can be compared against the experiment results (or the standard results) and the error percentage is calculated. As per the desired tolerance for the error, the simulations can be redesigned and re-ran to get accurate results.

Once the results are deemed acceptable, the big challenge is to interpret the results and process the output data so that they can be used in developing machine learning algorithms. The data generated by the simulation is in a very raw/crude form and could not be directly used to develop machine learning or response surface models. One of the major problem faced is that the data points are not evenly distributed over a time period. This is caused because of the different convergence techniques used by most numerical simulators. Each simulator has a pre-defined convergence limit which is guided by the minimum time-step provided to the simulator. The minimum time-step is defined to make sure the equations converge and doesn't give any errors or doesn't introduce any artifacts in the results. The initial time-step is chosen carefully according to the nature of equations used and depending on the physical process and the time scale. For example, a simulation with a fixed time-step would generate large number of data points and would also require more time to run. Whereas, using adaptive time-step, the simulator initially generate data points at very small time interval and as the equations begin to converge, it gradually increases the time-step and hence increasing the interval for data point generation. This leads to comparatively small number of data points as compared to the fixed time-step but even then the result might contain unnecessary amount of data points.

In this study, we have considered data obtained from a simulator for a temperature decline curve for the enhanced geothermal system(Asai, Panja et al. 2018). The temperature profile in geothermal system has been simulated or determined by analytical solution in case of relatively simple system(Wu, Zhang et al. 2014, Hadgu, Kalinina et al. 2016, Mudunuru, Karra et al. 2017, Asai, Panja et al. 2018). Uneven intervals are observed in simulated results especially in time series data. Due to the nature of numerical solution method, small time steps are required initially near t=0. On the other hand, larger time steps are used in the later time when the system is more stabilized. To demonstrate this numerical fact, produced water temperature consists of 9936 points from an enhanced geothermal system is shown in **Figure 1.**

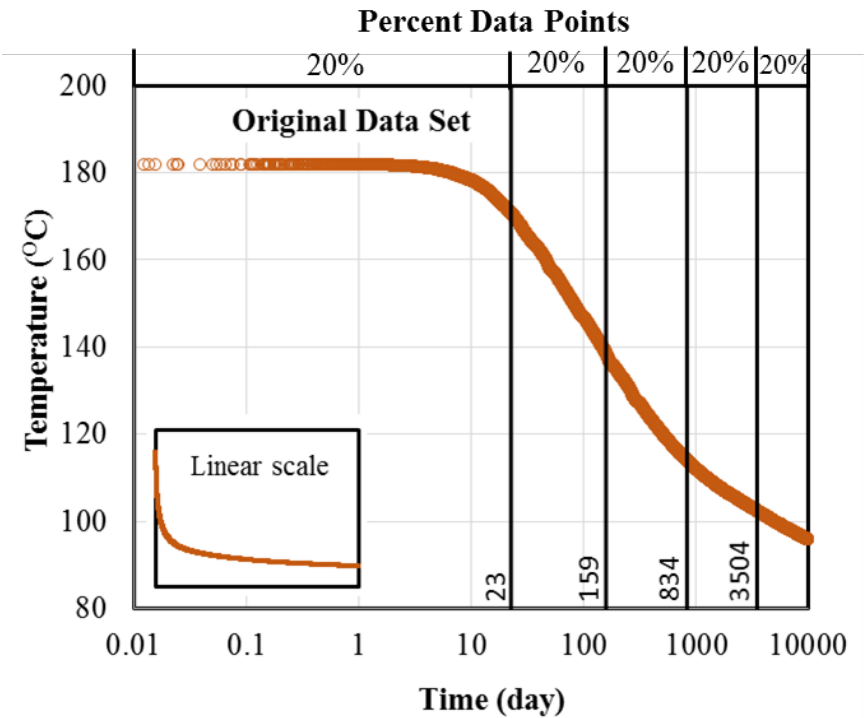

Figure 1: The data density along the X-axis

The logarithmic scale is used to expand the early time. The actual shape of the temperature profile in linear scale is also shown inside the figure. The X-axis i.e., time is divided into five groups such a way that each group contains 20% data. The first 20% of the data is from 0-23 days. Whereas last 20% of the data is from 6496 days (from 3504 to 10000 days). It is evident from this distribution that the data is highly dense towards initial time period. The density of data (number of points per unit time) is reduced towards the terminal time. In some instances, localized dense data is also observed due curvature of the profile. More data points are required to represent any curvature i.e., changes in slope along curve compared to fixed slope or straight-line portion of the curve.

This uneven distribution of data points over the entire time period, poses a problem in regression of a nonlinear mathematical equation as the data points are drastically skewed towards the beginning of the simulation. Thus, deeming the data points towards the end of simulation as the outliers and hence sometimes model cannot be fitted properly.

The study focuses on tackling such uneven distribution in the simulation results by implementing smart and novel techniques on preprocessing of the data so that it could be used in regression or curve fitting. To validate proposed data processing techniques, we considered the simulated data for the temperature decline curve in an enhanced geothermal reservoir generated through a commercial simulator.

The objective of this study is to reduce the total number of points to represent the entire curve to facilitate post processing of data such as curve fitting. Curve fitting using regression is highly dependent on number points as well as the local density of points. Various schemes are investigated to reduce the total number points and to obtain a better fit.

**Methodology**

Various steps involved in developing the protocol to reduce number of points in an uneven time series for regression are discussed here. All steps are summarized in a workflow for better comprehension in **Figure 2**.

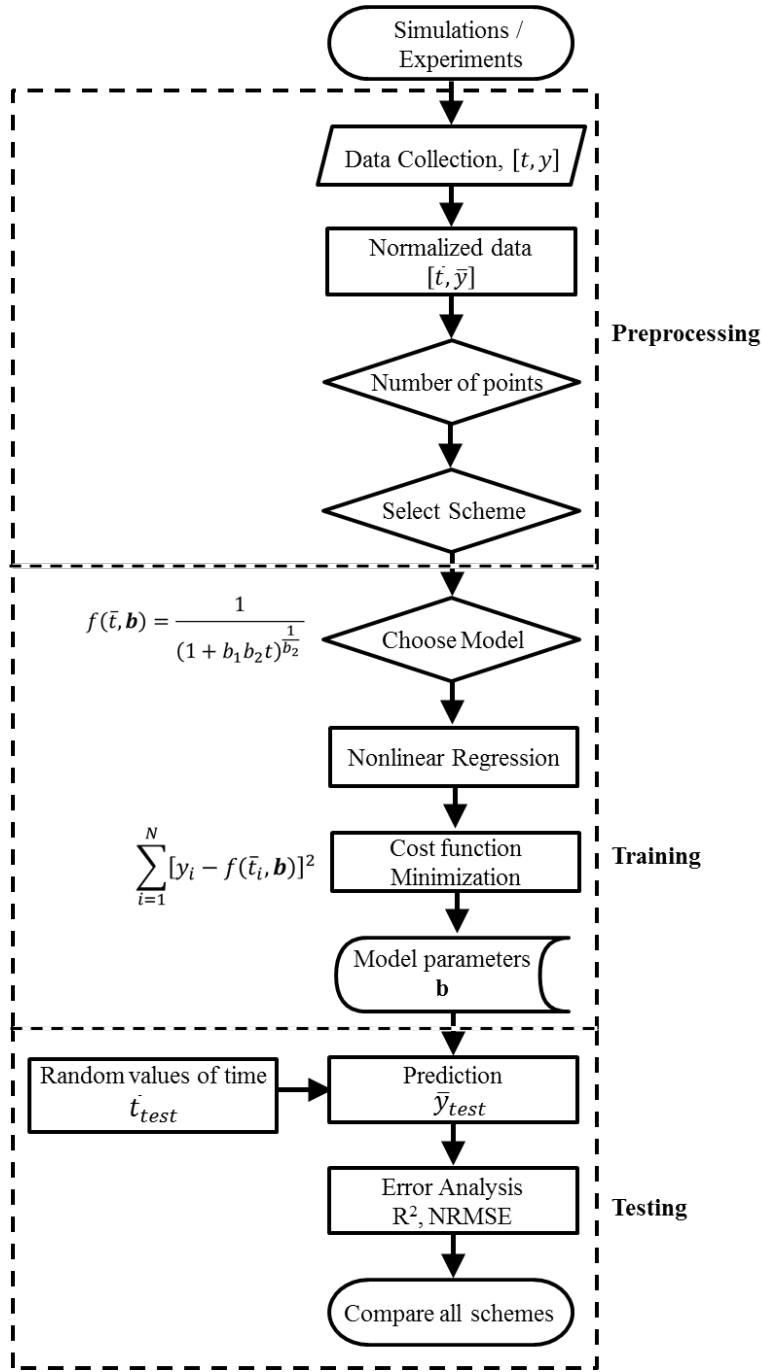

**Figure 2**: Workflow for data preprocessing, training and testing of fitted model

The workflow can be divided into three broad sections namely data preprocessing, training of mathematical model and testing of the fitted model. Individual components of the workflow such as normalization of data, followed by the various schemes and mathematical model for curve fitting are discussed in the next sections.

*Normalization*

It is observed in most of cases if not all that the ranges (minimum to maximum) and actual values of Y-axis and X-axis are not comparable in same scale. For example, in figure 1, the temperature (Y-axis) varies from 92 to 182 °C (range 90 °C) and time (X-axis) varies from 0 day to 10,000 days (range 10,000 days). In this case X-axis is more sensitive to regression compared to the Y-axis. To avoid this mathematical problem, it is advised to normalize the data to 0 to 1 for both axes. This will ensure the same ranges and actual values within the same minimum and maximum bracket. In the context of a geothermal system, the temperature and time are normalized as shown in **Equations 1** and **2**

$$\bar{T} = \frac{T - T_{min}}{T_{max} - T_{min}} \ldots\ldots\ldots\ldots\ldots\ldots\ldots\ldots\ldots\ldots\ldots\ldots. (1)$$

$$\bar{t} = \frac{t - t_{min}}{t_{max} - t_{min}} \ldots\ldots\ldots\ldots\ldots\ldots\ldots\ldots\ldots\ldots\ldots\ldots. (2)$$

Using the above formulae, the minimum temperature i.e., 92 °C becomes zero and maximum temperature 182 °C becomes 1. Similarly, minimum and maximum times become 0 and 1 too. Using the normalized data, six schemes are tried in this study to select certain number of points from total points of 9936 (see Figure 1) as shown in the **Table 1**.

**Table 1**: Various schemes for selecting points to reduce the number of points

| Label | Method |
|---|---|
| **Scheme 1** | Entire Original Data set |
| **Scheme 2** | Equal division of X-axis |
| **Scheme 3** | Equal division of Y-axis |
| **Scheme 4** | Equal division along curve |
| **Scheme 5** | Constraint in deflection |
| **Scheme 6** | Mixed of schemes 3 and 4 |

Following the normalized temperature and time from geothermal system, all schemes are discussed in the next sections.

*Scheme 1: Entire data set*

In this scheme, data set is kept unchanged i.e., entire data of 9939 points are chosen for regression. This is base case for comparison with other schemes to establish the effectiveness of the data preprocessing. All points are plotted in **Figure 3**.

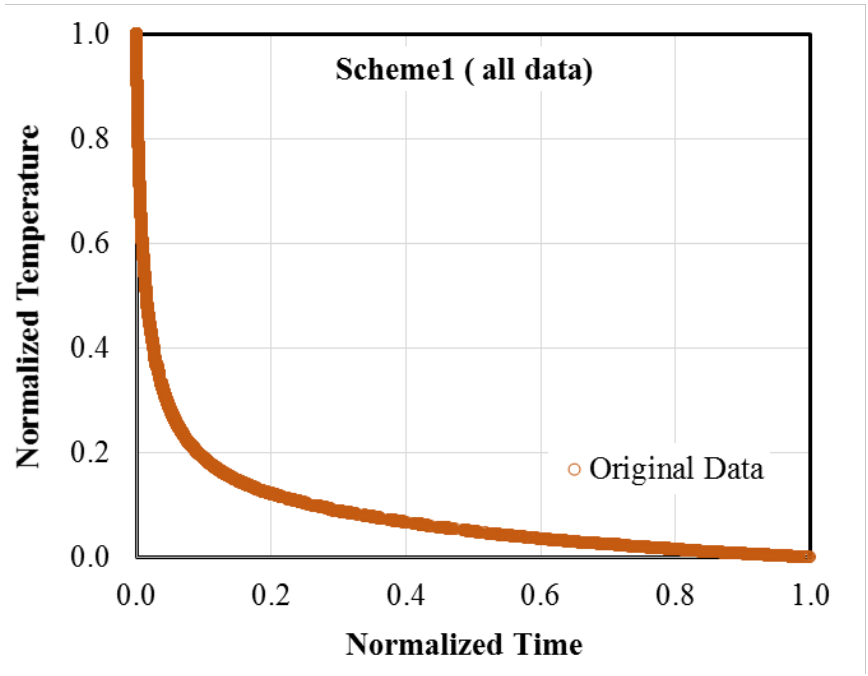

**Figure 3:** Scheme 1 where no changes are made and the entire data set is selected.

## Scheme 2: Equal division of X-axis

In this scheme, the temperatures are selected such a way that they have equal interval i.e., the entire range of data in the X-axis is divided equally. Spacing between two consecutive points in this scheme is calculated as given in equation 3

$$\Delta X = \Delta t = \frac{t_{max} - t_{min}}{N - 1} = \frac{t_{max}}{N - 1}, \quad where \ t_{min} = 0 \ldots\ldots\ldots\ldots\ldots\ldots. (3)$$

Any data point in this scheme is calculated by equation 4

$$t_i = t_{min} + \Delta T \ (i - 1), \quad i = 1,2,3..N, \qquad \ldots\ldots\ldots\ldots\ldots\ldots\ldots. (4)$$

To display the position of the points on the curve, about 1% of total data (100 points from 9939 points) are chosen as shown in **Figure 4**. Details discussion of sensitivity of number of points on curve fitting is provided in results section. 1% data is considered as case 1 in the sensitivity study and it is used for all figures showing different schemes for demonstration.

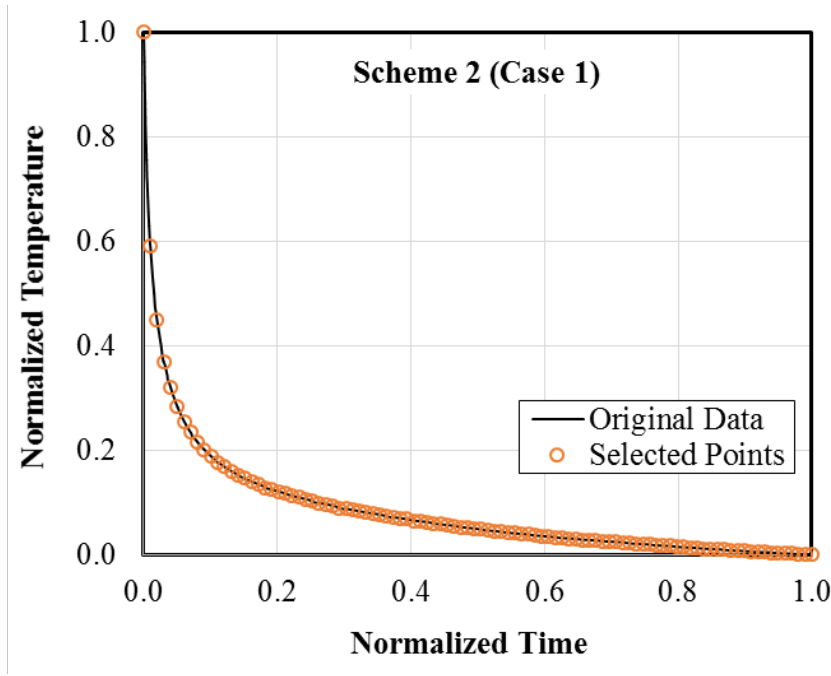

**Figure 4**: Scheme 2 where points are located with equal interval in X-axis i.e., time

It is evident from the figure 4 that the distance between two consecutive points varies depending on the curvature or slope of the curve with respect to time $\left(\frac{\Delta T}{\Delta t}\right)$. More points are located at lower slope section such as the flat potion of the curve compared to higher slope section. At higher slope $\left(\frac{\Delta T}{\Delta t}\right)$, same amount of change in time ($\Delta t$) has more changes in temperature($\Delta T$), thus larger distance between two points is observed. Technically point density can be defined as the number of points in a fixed one dimensional length (may be curve or straight line). Same point density at any location on the curve can be found for the curve with constant slope i.e., for linear equation.

### *Scheme 3: Equal division of Y-axis*

Like the equal division of X-axis, the entire range of Y-axis can be divided into equal interval as shown in **Figure 5**.

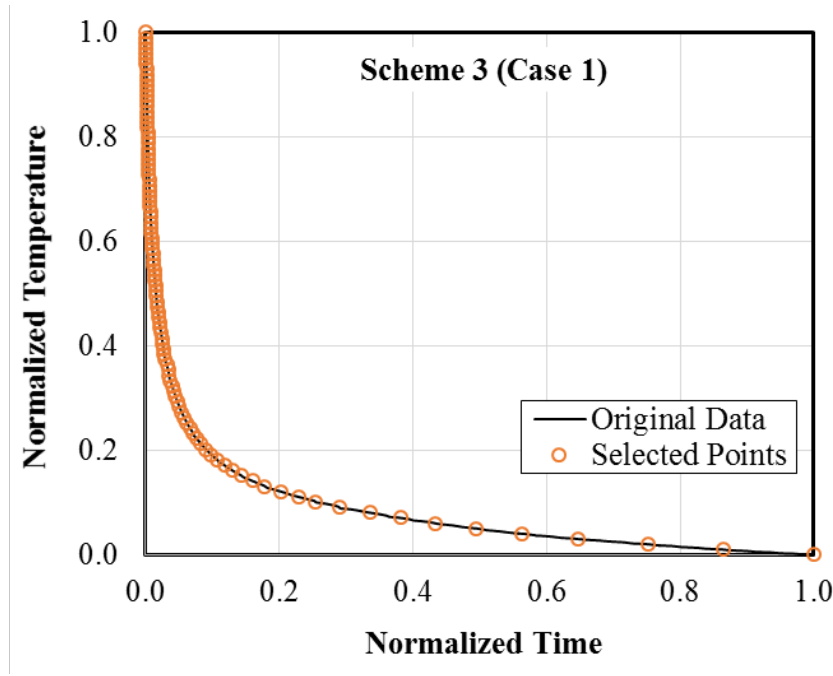

**Figure 5**: Scheme 3 where points are located with equal interval in Y-axis i.e., temperature Spacing in this scheme is calculated as given in **equation 5**

$$\Delta Y = \Delta T = \frac{T_{max} - T_{min}}{N - 1} \quad \dots \dots \dots \dots \dots \dots \dots \dots \dots \dots \dots \dots \dots \dots \dots (5)$$

In the case of a geothermal system, temperature starts initially at maximum value and reduces towards the minimum as time goes. Any data point in this scheme is calculated by **equation 6**

$$T_i = T_{max} - \Delta T\,(i - 1), \quad i = 1,2,3 \dots N, \dots \dots \dots \dots \dots \dots \dots \dots \dots \dots \dots \dots (6)$$

Location of higher point density in the curve is opposite to the previous scheme. A higher point density is found in the higher slope section. At higher slope $\left(\frac{\Delta T}{\Delta t}\right)$ region, for same change in temperature($\Delta T$), change in time($\Delta t$) in time is less which ensures more points in the region.

*Scheme 4: Equal division along curve*

In this scheme, the entire curve is divided into equal pieces along the trajectory of the curve. First task in this scheme is to calculate the length of the entire curve. Next, the total length is divided into specified number to get desired number of points with equal spacing along the curve. The distance between two consecutive points ($\Delta L$) is calculated using equation 7 and demonstrated in **Figure 6**.

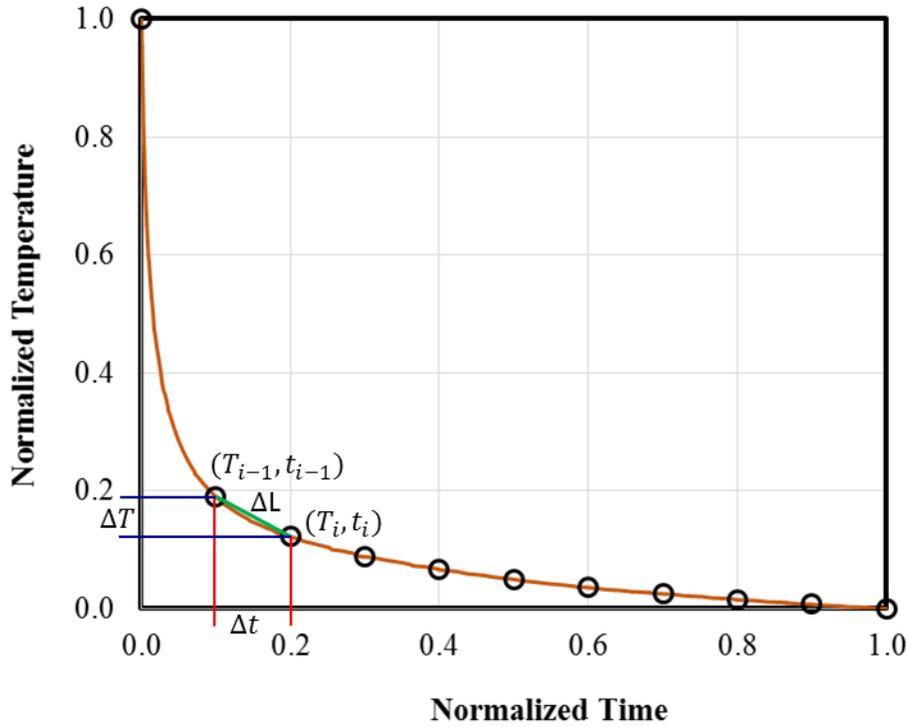

**Figure 6:** The demonstration of intervals and distance between two consecutive points

$$\Delta L_{i,i-1} = \sqrt{(T_i - T_{i-1})^2 + (t_i - t_{i-1})^2} \quad \ldots \ldots (7)$$

To calculate the length of the entire curve, distances between two consecutive points (as calculated using equation 7) are added together as shown in equation 11

$$L = \sum_{i=2}^{N} \Delta L_{i,i-1} = \sum_{i=2}^{N} \sqrt{(T_i - T_{i-1})^2 + (t_i - t_{i-1})^2} \quad \ldots \ldots (8)$$

If the entire length along the curve is divided into N equal spacing, then the interval is calculated as

$$\Delta L = \frac{L}{N-1}, \quad \ldots \ldots (9)$$

Placing points with ΔL interval along curve is calculated as

$$t_i = t_{i-1} + \sqrt{\Delta L^2 - (T(t_i) - T_{i-1})^2} \quad \ldots \ldots (10)$$

As shown in equation 10 that the method to find out the points with ΔL interval with previous point is an iterative method. The current time step ($t_i$) is assumed first, the temperature ($T_i$) is interpolated from the original data set. Then, using calculation shown in equation 10, current time step ($t_i$) is calculated again. If the assumed and calculated time match, then current time step is accepted and proceed for next time step. The points calculated in this way are shown in **Figure 7**.

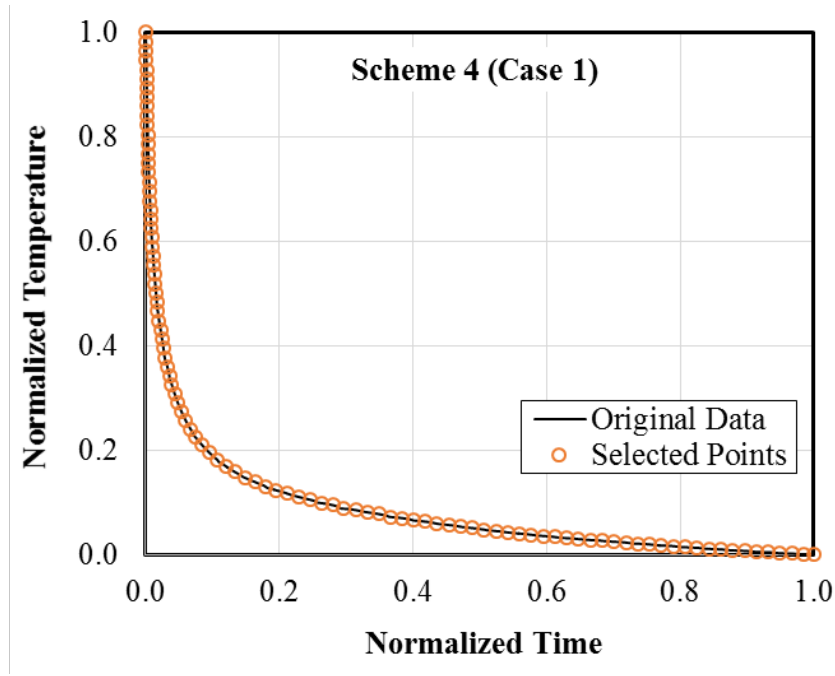

**Figure 7**: Scheme 4 where points are chosen based on equal division of the length of the curve

In this scheme, it is clear that each section of the curve has same point density irrespective of the slope of the curve.

*Scheme 5: Constraint in deflection*

In the previous schemes, especially in the schemes 2 to 4, the fixed number of total points is selected based on their criteria. Selection of total number of points is totally knowledge based. The total number of points should be sufficient to capture all the features of the curvature. Scheme 5 is formulated to ensure that the curved sections with sharp changes in slope are considered during regression. In this scheme, sum of change in slopes (in terms of angle) in successive points or deflection in the curve is set as a criterion. The calculation of angle in this scheme is explained in **Figure 8**.

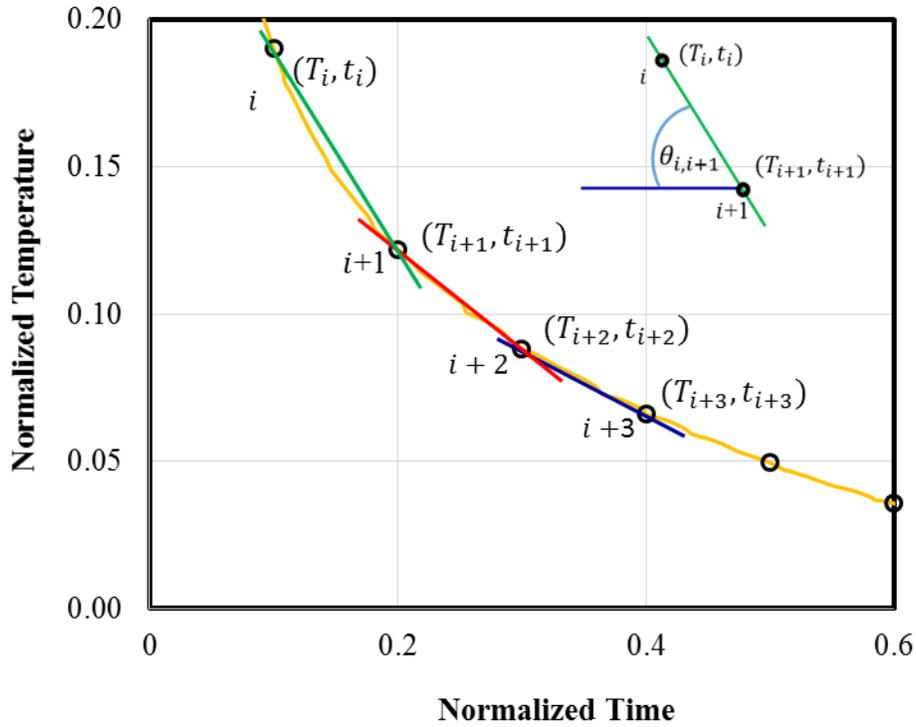

**Figure 8**: Deflection based scheme where points are chosen based on user specified angle of deflection

The slope and corresponding angle between two consecutive points are calculated as described in equations 11 and 12

$$S_{i,i+1} = \left(\frac{T_{i+1} - T_i}{t_{i+1} - t_i}\right) \quad \ldots\ldots\ldots\ldots\ldots\ldots\ldots (11)$$

$$\theta_{i,i+1} = \tan^{-1}(S_{i,i+1}) \quad \ldots\ldots\ldots\ldots\ldots\ldots\ldots (12)$$

To calculate the total deflection from the (i+1)th point, angles between next few points such as $\theta_{i,i+1}$, $\theta_{i+1, i+2}$, $\theta_{i+2, i+3}$ should be known. The total change in angle from point i to next k points is calculated as

$$\Delta\theta_{i,k} = (\theta_{i+1,i+2} - \theta_{i,i+1}) + (\theta_{i+2,i+3} - \theta_{i+1,i+2}) + \cdots + (\theta_{i+k,i+k+1} - \theta_{i+k-,i+k})$$

$$= \sum_{j=1}^{k}(\theta_{i+j,i+j+1} - \theta_{i+j-1,i+j}) \quad \ldots\ldots\ldots\ldots\ldots\ldots\ldots\ldots\ldots\ldots\ldots (13)$$

Then the percentage change in slope with respect to angle between i and i+1 points is calculated as

$$\Delta\theta_i = \left|\frac{\theta_{i+1,i+2} - \theta_{i,i+1}}{\Delta\theta_{i,k}}\right| \times 100 \quad \ldots\ldots\ldots\ldots\ldots\ldots\ldots\ldots\ldots (14)$$

Points are added until the calculated $\Delta\theta_i$ is met the set criteria of deflection. If the calculated $\Delta\theta_i$ according to equation 14 is acceptable for a given value (say 75%), then the $k^{th}$ point after $i^{th}$ point is acceptable as the next i+1 point in the scheme. The points calculated this way with 75% of deflection tolerance are shown in **Figure 9**

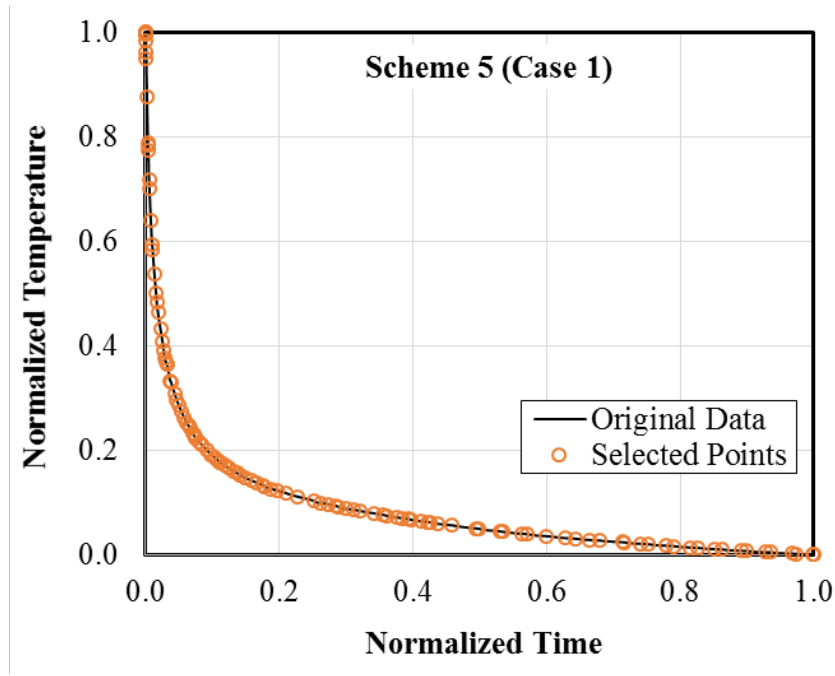

**Figure 9**: Scheme 5 where deflection is chosen as criteria to select representative points

This scheme ensures that no parts of curves with sharp slope change are ignored for the regression. Intervals in the curve are irregular depending on the localized slope. In original data, sudden changes of slope are observed in many places and it causes some close spacing of points. This cause the different weightage of different section of the curve in the cost function.

### *Scheme 6: Mix of schemes 4 and 5*

To reduce the number of total points further, the scheme 4 and 5 are mixed together. The scheme 6 is same as scheme 5 except that the points generated in scheme 4 are considered instead of original data set (scheme 1). In scheme 4, equally spaced points along curves are generated. A few neighboring points might not have sharp changes in slopes. Using this scheme those neighboring point could be merged together. The points are significantly reduced compared to figures 7 and 8 as shown in **Figure 10**.

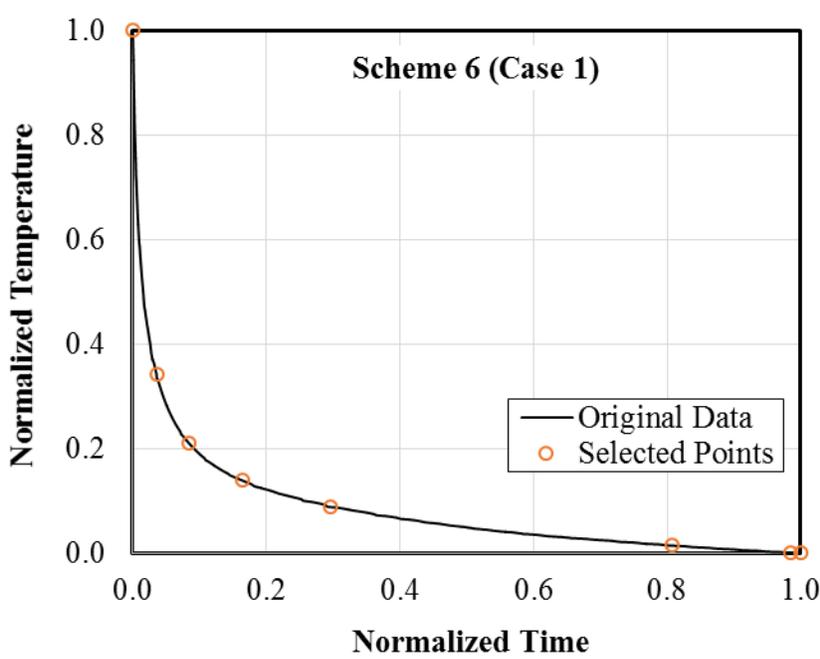

**Figure 10**: Selection of points based on 75% cumulative deflection change on the equally spacing points along curve

It is noticed that four points out of seven points are located on the deflection section. On the other hand, few points are located on the straight-line sections.

*Regression*

After choosing points from experimental data based on the criteria set by each scheme, a mathematical model is fitted. Reviewing several time series forecasting models which are available now, De Gooijer and Hyndman (De Gooijer and Hyndman 2006) classified them into eight categories namely (i) exponential smoothing(Muth 1960, Gardner 1985, Snyder 1985), (ii) Autoregressive Integrated Moving Average (ARIMA) (Yule 1927, Box and Jenkins 1970), (iii) seasonal models(Dagum 1982, Huyot, Chiu et al. 1986), (iv) state space and structural models and the Kalman filter (Kalman 1960, Schweppe 1965, Shumway and Stoffer 1982), (v) nonlinear models (Volterra 1930, Wiener 1958), (vi) long-range dependence models, e.g. the family of Autoregressive Fractionally Integrated Moving Average (ARFIMA) models(Ray 1993, Ray 1993), (vii) Autoregressive Conditional Heteroscedastic/Generalized Autoregressive Conditional Heteroscedastic (ARCH/GARCH) models (Engle 1982, Taylor 1987, Bollerslev, Engle et al. 1994) and (viii) count data forecasting(Croston 1972, Willemain, Smart et al. 1994). In this study, a nonlinear function, $f(\bar{t}, \boldsymbol{b})$ is chosen for regression as shown in **Equation 15**

$$\bar{T} = f(\bar{t}, \boldsymbol{b}) = \frac{\bar{T}_0}{(1 + b_1 b_2 t)^{\frac{1}{b_2}}} \ldots \ldots \ldots \ldots \ldots \ldots \ldots (15)$$

$$\text{where,} \quad \bar{T}_0 = \frac{T(t_{min}) - T_{min}}{T_{max} - T_{min}} \ldots \ldots \ldots \ldots \ldots \ldots \ldots (16)$$

The above equation is originally applied by Arps (Arps 1945) for decline in oil rate. In case of normalized data (0 to 1), the ($\bar{T}_0$) becomes one, therefore only parameters required to determine using regression are $b_1$ and $b_2$. In the curve fitting method, a cost function which is generally the sum of the errors or square of the sum of errors is minimized. An optimization routine '*nlinfit*' in Matlab (Mathworks Inc.) is used where the cost function is given by the **Equation 17**

$$\text{cost function} = \sum_{i=1}^{N}[y_i - f(\bar{t}_i, \boldsymbol{b})]^2 \ldots \ldots \ldots \ldots \ldots \ldots \ldots \ldots \ldots (17)$$

The $y_i$ is the normalized temperature from the experiments or simulations and $f(\bar{t}, \boldsymbol{b})$ is the normalized temperature predicted by the model for same normalized time. Total N points or observations are used for the fitting. The quality of the fitted function is evaluated by various statistical measurements such as coefficient of determination ($R^2$), error (e), mean square error (MSE), normalized root mean square error (NRMSE) etc. which are discussed in the Appendix A.

**Results and Discussions**

As described in the workflow (Figure 2), the results are analyzed for both training and test data sets. Temperature profiles and errors in predictability are discussed here for scheme 1 to 7 for various scenario.

*Training of Models*

In this study, temperature versus time data are simulated from an enhanced geothermal system. Details of the simulation of geothermal system is not relevant to this topic, therefore only the data from the simulation is presented here. Total 9939 points are collected to represent 0 to 10,000 days. To study the sensitivity of the chosen number of data points on the fitted curve, 8 cases are investigated by varying the number of selected points (from 1 to 20 %) as given in **Table 2**.

**Table 2**: Data utilization in each scheme for various cases. (100% data is used for scheme 1)

| Case | Percentage of total data | | Number of data | |
|---|---|---|---|---|
| | Scheme 2 to 5 | Scheme 6 | Scheme 2 to 5 | Scheme 6 |
| Case 1 | 1 | 0.08 | 100 | 8 |
| Case 2 | 2 | 0.3 | 199 | 30 |
| Case 3 | 3 | 1.1 | 299 | 103 |
| Case 4 | 4 | 1.8 | 398 | 177 |
| Case 5 | 5 | 2.2 | 497 | 214 |
| Case 6 | 10 | 3.8 | 994 | 375 |
| Case 7 | 15 | 5.4 | 1491 | 533 |
| Case 8 | 20 | 6.9 | 1988 | 690 |

In scheme 1, 100% data is used, therefore only single case is applied for scheme 1 and not shown in the above table. For schemes 2 to 5, 1 to 20% of total data are used to estimate the model parameters in equation 15. In scheme 6, number of data points (0.08% to 6.9%) are less than the data points used in scheme 4 because of the nature of the scheme. In scheme 5, choosing any number of points is not straightforward like scheme 2 to 4. In this scheme, the constraint in deflection (angle of curvature) is set as criteria instead of number of points. Therefore, it is a trial and error method to choose an angle such a way that the scheme will have certain number of points. Angles for scheme 5 are 75, 15.5, 2.6, 1.07, 0.70, 0.235, 0.127 and 0.0785 for cases 1 to 8 respectively.

Model provided in equation 15 is trained for all the cases of various scheme. The model parameters are enlisted in **Table 3**.

**Table 3**: Model parameters $b_1$, $b_2$ after regressions of models (equation 15) from various schemes for different cases

| Model Parameter | Case | Scheme 2 | Scheme 3 | Scheme 4 | Scheme 5 | Scheme 6 |
|---|---|---|---|---|---|---|
| b1 | Case 1 | 1.20 | 1.47 | 1.34 | 1.37 | 1.19 |
|    | Case 2 | 1.21 | 1.47 | 1.34 | 1.36 | 1.33 |
|    | Case 3 | 1.21 | 1.47 | 1.35 | 1.37 | 1.34 |
|    | Case 4 | 1.21 | 1.47 | 1.35 | 1.37 | 1.34 |
|    | Case 5 | 1.20 | 1.47 | 1.34 | 1.38 | 1.35 |
|    | Case 6 | 1.21 | 1.47 | 1.33 | 1.37 | 1.37 |
|    | Case 7 | 1.21 | 1.47 | 1.30 | 1.37 | 1.28 |
|    | Case 8 | 1.21 | 1.47 | 1.28 | 1.37 | 1.22 |
| b2 | Case 1 | 59.1 | 75.5 | 71.5 | 70.2 | 57.2 |
|    | Case 2 | 59.5 | 75.5 | 71.7 | 69.7 | 68.4 |
|    | Case 3 | 59.5 | 75.6 | 71.8 | 70.5 | 69.8 |
|    | Case 4 | 59.5 | 75.6 | 71.7 | 70.1 | 70.2 |
|    | Case 5 | 59.4 | 75.6 | 71.7 | 70.3 | 70.2 |
|    | Case 6 | 59.4 | 75.6 | 71.4 | 68.9 | 71.0 |
|    | Case 7 | 59.4 | 75.6 | 70.3 | 68.6 | 67.7 |
|    | Case 8 | 59.4 | 75.6 | 69.1 | 68.6 | 65.3 |

Model parameters b1 and b2 for scheme 1 are 1.40 and 72.3 respectively. It is evident that in the most of the schemes, parameters do not vary a lot for different cases i.e., the parameters are independent of the number of points chosen for training. Only noticeable variations in parameters are observed for different cases in scheme 6. Significantly low number of points compared other schemes may be the reason behind these results in scheme 6. On the other hand, the parameters vary for different scheme in fixed case. The fitted model from various scheme and the original observation are compared in **Figure 11** for case 1 (1% data).

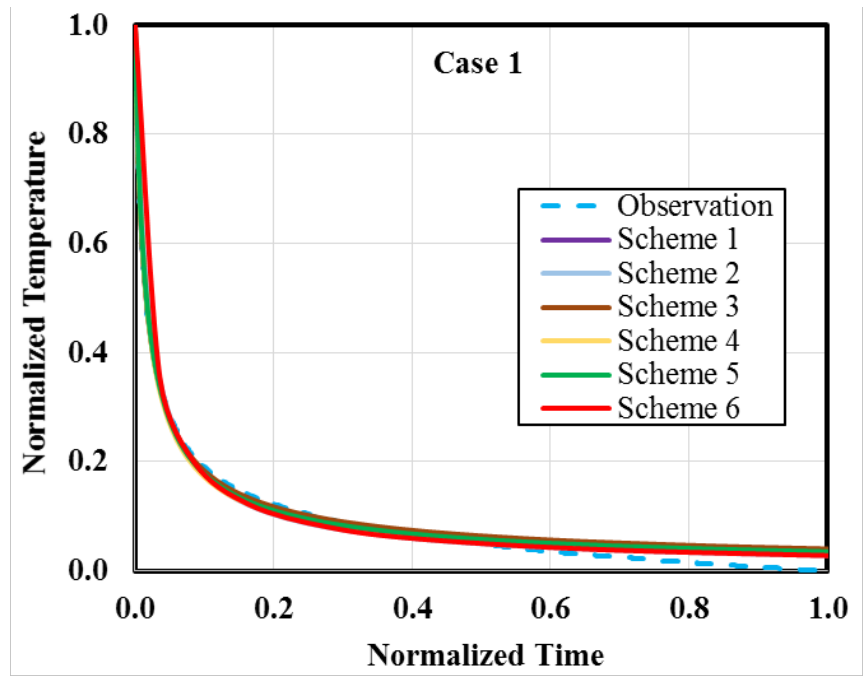

**Figure 11**: Fitted functions and using schemes 1 to 6 for case 1

Although the values of model parameters, b1 and b2 (see Table 2) vary from scheme to scheme, a good agreement is observed between fitted model and experimental data. This indicates that the combination of values b1 and b2 in the model (equation 15) works well in prediction. To visualize the difference between fitted model and experimental data, error is plotted in **Figure 12**.

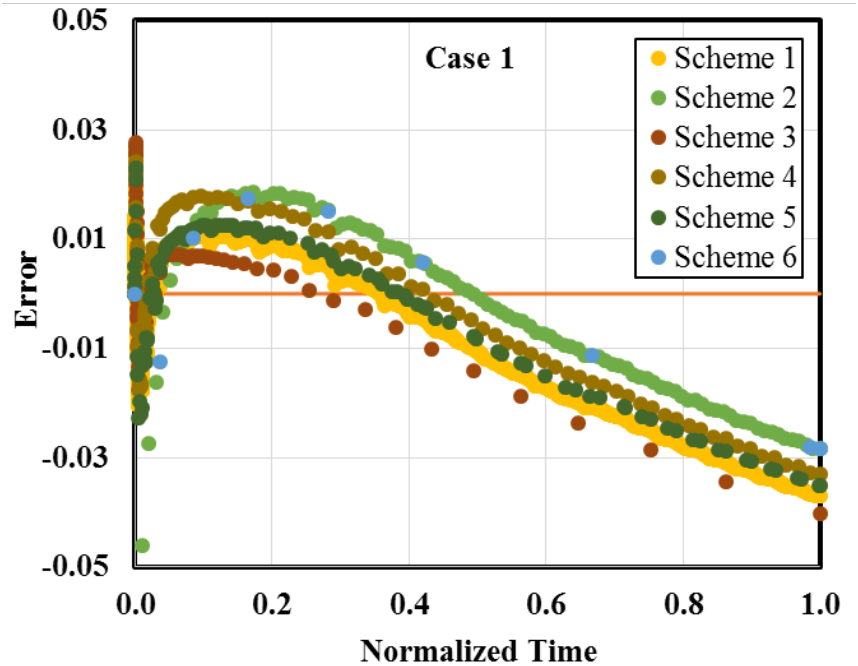

**Figure 12**: Errors using case 1 for schemes 1 to 6

Because of the fact that data was normalized, all points fall between 0 and 1, therefore, the errors in the range of -0.05 to 0.05 are not very significant. Errors vary with time for all schemes. The highest errors are observed at the end of the curve and at the point of deflection of the curve. It means that the fitted model had difficulty to represent the curvature section of the data. Scheme 3 had low errors initially but it started deviating in the later time. Surprisingly, scheme 1 where 100% data are used had significant errors. Schemes 2, 4, 5 and 6 showed better fit compared to other schemes by showing overall less error for entire time period. Although, scheme 2 had highest error in the initial portion. Schemes 4 and 5 are possibly the best fit where low error is observed in the curvature section. This is because of the nature of the selection criteria of points in schemes 4 and 5. All points are equally spaced on the curve providing equal weightage to each section of curve in curve fitting. On the other hand, in scheme 5, the curve section had sufficient point for more weightage in the curve fitting, in other words, the cost function defined in equation 17 is more influenced by this section.

Overall fitness of the model for various cases in different schemes is measured by the combined error such as coefficient of determination ($R^2$) and normalized root mean square error (NRMSE) as shown in **Figure 13**.

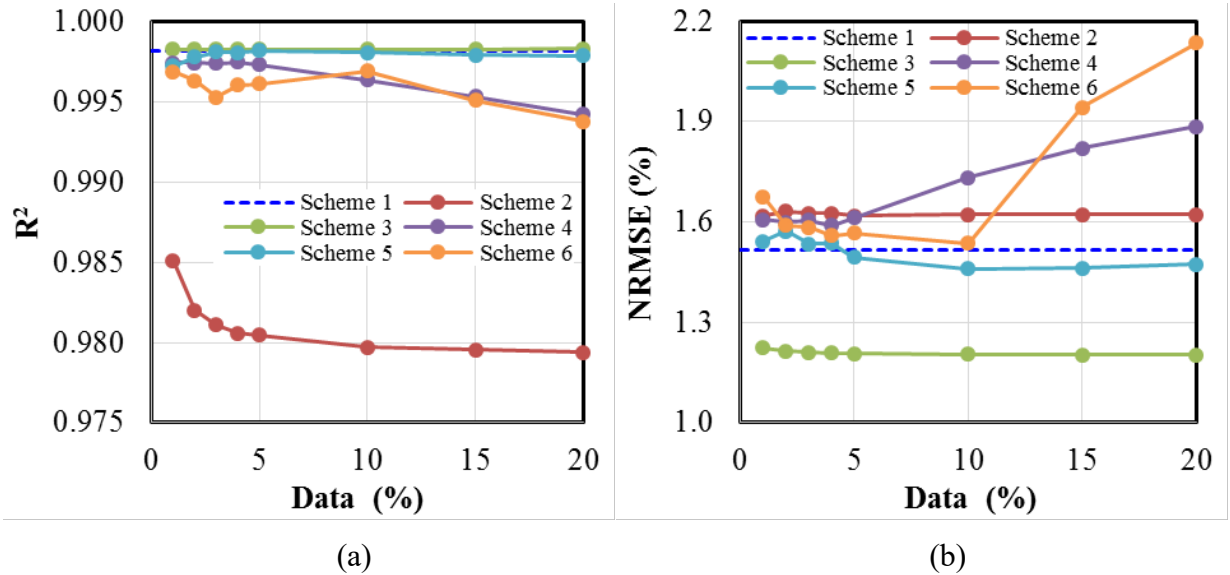

**Figure 13**: Error analysis of fitted functions for various scheme and different cases (a) Coefficient of determination ($R^2$) (b) Normalized Root Mean Square Error (NRMSE)

Different percentage in the X-axis in the above figures are the different cases as shown in table 2. Because scheme 1 has only one case (100% data), one $R^2$ and one NRMSE values are calculated which are shown by dotted blue lines in the figures. Although the coefficient of determinations ($R^2$) of fitted curves for all schemes except scheme 2 in figure 13(a) are high which an indication of good fit, the NRMSEs are also in the higher side for schemes 4 and 6 which is an indication of bad fitting.

Considering the results from figures 12 and 13, schemes 4 and 5 are the best choice to select points for regression. Scheme 2 could be another choice but the initial deviations (see figure 12) make it unsuitable.

## Testing of Models

Fitness of a model is not always guaranteed by the error analysis from training data. Test data set which is essentially randomly picked points within the range of study is required to check the predictability of the model. In testing of various schemes, model parameters (b1 and b2) obtained from case 5 where 5% data were used in regression are chosen without any strict technical reasons. As shown in the workflow (figure 2), one hundred random values of time are chosen in the range of 0 to 1 for testing of the model. The predicted values from scheme 1 to scheme 6 are compared with the experimental/simulated data in **Figure 14**.

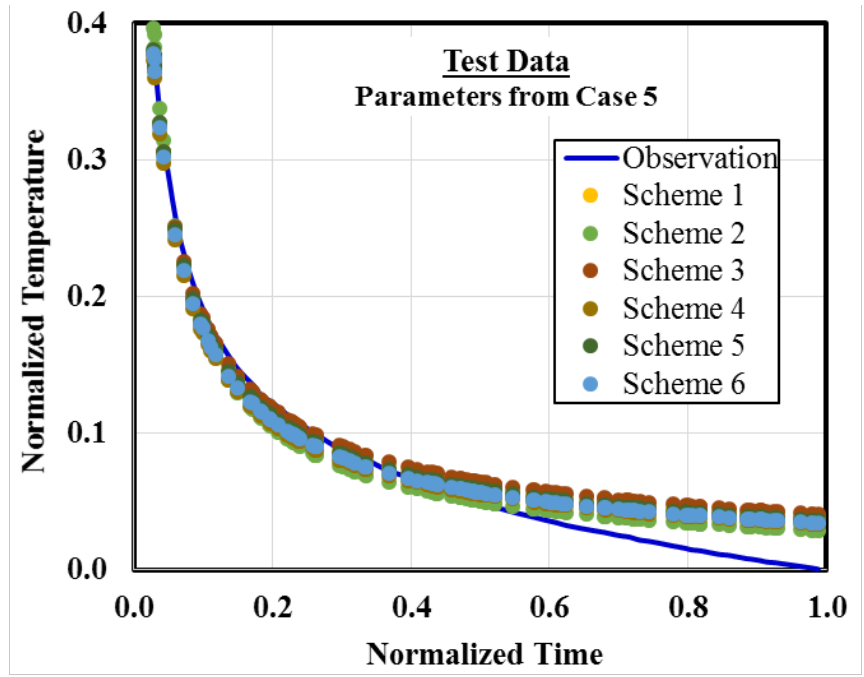

**Figure 14:** Testing of various scheme for model fitness using randomly chosen 100 values of time

The predicted values from different schemes have discernible differences. All schemes predicted well close to the observed values until the midway, after that values are underestimated. The highest differences are observed at the end. To differentiate each scheme, the error plots are shown in **Figure 15**.

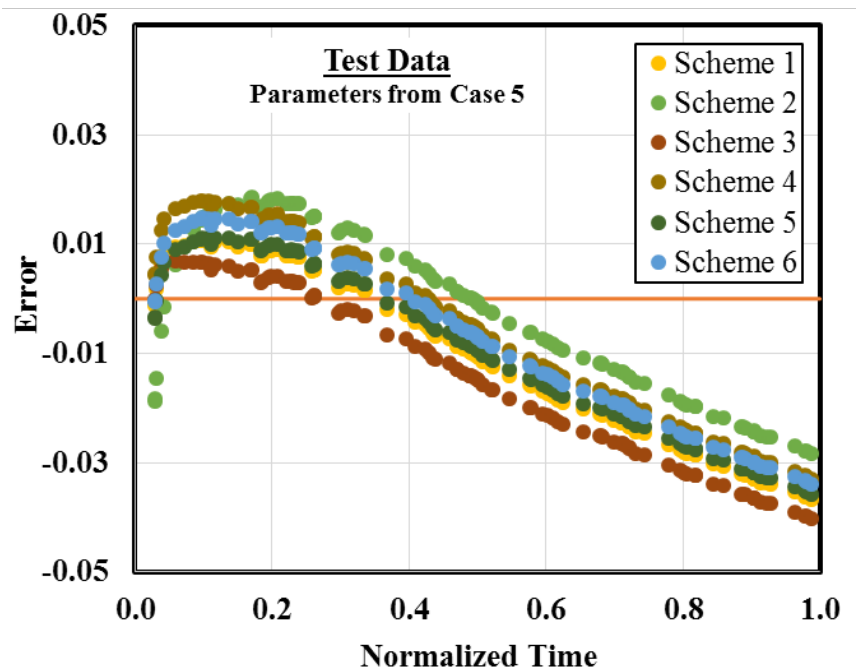

**Figure 15:** Error in the predictions from scheme 1 to 6 for testing data

Like training set, scheme 2 has the highest error in the initial time period but errors are low in the later time. Schemes 4, 5 and 6 have the low variations in error throughout the entire time period. This can be evident from the quantitative error analysis ($R^2$ and NRMSE) as shown in **Figure 16**.

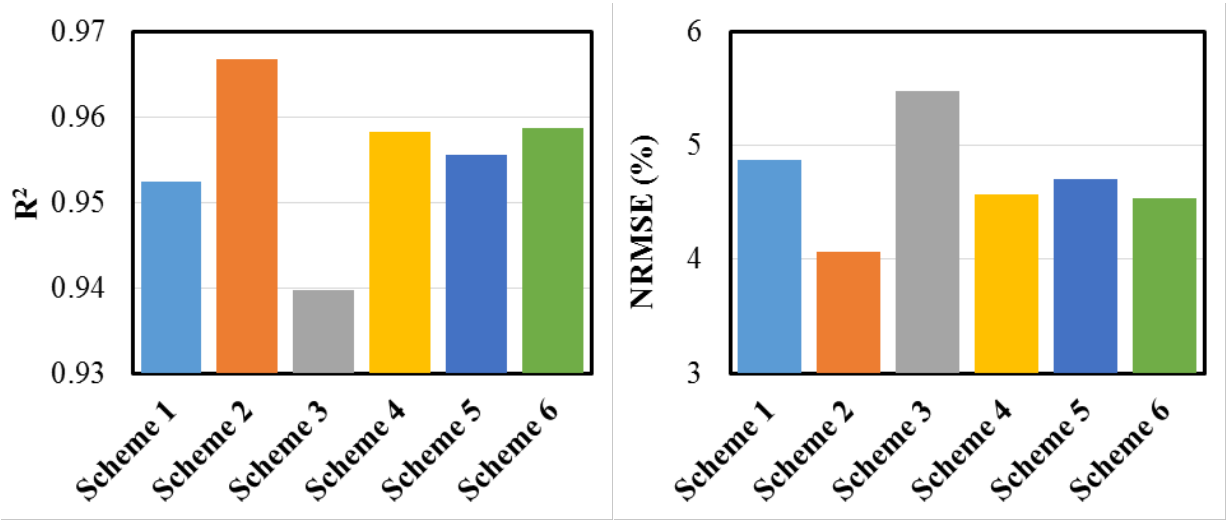

**Figure 16:** Quantitative error analysis of various scheme for testing set (a) Coefficient of determination (R2) (b) Normalized Root Mean Square Error (NRMSE)

The scheme 2 has the highest R2 and lowest NRMSE which indicate the best fit among all scheme, however, due the higher initial error (figure 16), this scheme may not be the best to predict for entire time period. Scheme 3 has the lowest $R^2$ and the highest NRMSE values which make it least suited scheme among all. Like training data set, schemes 4, 5 and 6 remain the best schemes where higher $R^2$ and lower NRMSE are observed.

**Conclusions**

Despite the large availability of data points, fitted functions often fail to represent all these points due to differences in data density. 1% to 20% data points are selected using 6 different selecting criteria (schemes 1 to 6). It is shown that even 1% worth of data points is as good as the entire data set for regression as long as the proper scheme to select points from original data set is chosen. Scheme 4 (where distance between two consecutive points is fixed) and 5 (where angle of deflection is the criteria) are the most efficient schemes that provide a better fit of the mathematical model for any given number of points. Test data set which is chosen randomly confirms the robustness of these schemes. This study helps reduce the number of data points necessary during regression and improve the fit of any model when the data points are not evenly distributed.

**Conflicts of Interest**

The authors declare that there is no conflict of interest regarding the publication of this paper.

**Funding Statement**

Funding for this project was provided by Department of Chemical Engineering, University of Utah, Utah, USA.

**Acknowledgements**

The authors gratefully acknowledge the academic license to CMG products from Computer Modeling Group, Calgary, Canada.

**References**

Alom, M. S., M. Tamim and M. M. Rahman (2017). "Decline curve analysis using rate normalized pseudo-cumulative function in a boundary dominated gas reservoir." Journal of Petroleum Science and Engineering **150**: 30-42.

Arps, J. J. (1945). "Analysis of Decline Curves, SPE-945228-G." Transactions of the AIME **160**(01).
Asai, P., P. Panja, J. McLennan and J. Moore (2018). "Performance evaluation of enhanced geothermal system (EGS): Surrogate models, sensitivity study and ranking key parameters." Renewable Energy **122**: 184-195.
Beylkin, G., J. Garcke and M. J. Mohlenkamp (2009). "Multivariate Regression and Machine Learning with Sums of Separable Functions." SIAM Journal on Scientific Computing **31**(3): 1840-1857.
Bollerslev, T., R. F. Engle and D. B. Nelson (1994). Chapter 49 Arch models. Handbook of Econometrics, Elsevier. **4:** 2959-3038.
Box, G. E. P. and D. W. Behnken (1960). "Some New Three Level Designs for the Study of Quantitative Variables." Technometrics **2**(4): 455-475.
Box, G. E. P. and G. M. Jenkins (1970). Time series analysis: forecasting and control, Holden-Day.
Croston, J. D. (1972). "Forecasting and Stock Control for Intermittent Demands." Operational Research Quarterly (1970-1977) **23**(3): 289-303.
Dagum, E. B. (1982). "Revisions of time varying seasonal filters." Journal of Forecasting **1**(2): 173-187.
De Gooijer, J. G. and R. J. Hyndman (2006). "25 years of time series forecasting." International Journal of Forecasting **22**(3): 443-473.
Engle, R. F. (1982). "Autoregressive Conditional Heteroscedasticity with Estimates of the Variance of United Kingdom Inflation." Econometrica **50**(4): 987-1007.
Gardner, E. S. (1985). "Exponential smoothing: The state of the art." Journal of Forecasting **4**(1): 1-28.
Hadgu, T., E. Kalinina and T. S. Lowry (2016). "Modeling of heat extraction from variably fractured porous media in Enhanced Geothermal Systems." Geothermics **61**: 75-85.
Huyot, G., K. Chiu, J. Higginson and N. Gait (1986). "Analysis of Revisions in the Seasonal Adjustment of Data Using X-11-Arima Model-Based filters." International Journal of Forecasting **2**(2): 217-229.
Kalman, R. E. (1960). "A New Approach to Linear Filtering and Prediction Problems." Journal of Basic Engineering **82**(1): 35-45.
Kamari, A., A. H. Mohammadi, M. Lee and A. Bahadori (2017). "Decline curve based models for predicting natural gas well performance." Petroleum **3**(2): 242-248.
Mudunuru, M. K., S. Karra, D. R. Harp, G. D. Guthrie and H. S. Viswanathan (2017). "Regression-based reduced-order models to predict transient thermal output for enhanced geothermal systems." Geothermics **70**: 192-205.
Muth, J. F. (1960). "Optimal Properties of Exponentially Weighted Forecasts." Journal of the American Statistical Association **55**(290): 299-306.
Okouma Mangha, V., D. Ilk, T. A. Blasingame, D. Symmons and N. Hosseinpour-zonoozi (2012). Practical Considerations for Decline Curve Analysis in Unconventional Reservoirs - Application of Recently Developed Rate-Time Relations. SPE Hydrocarbon Economics and Evaluation Symposium. Calgary, Alberta, Canada, Society of Petroleum Engineers.
Ray, B. K. (1993). "Long-range forecasting of IBM product revenues using a seasonal fractionally differenced ARMA model." International Journal of Forecasting **9**(2): 255-269.
Ray, B. K. (1993). "MODELING LONG-MEMORY PROCESSES FOR OPTIMAL LONG-RANGE PREDICTION." Journal of Time Series Analysis **14**(5): 511-525.
Robinson, S. (2004). Simulation: The Practice of Model Development and Use. Chichester, England, John Wiley & Sons Ltd.
Schweppe, F. (1965). "Evaluation of likelihood functions for Gaussian signals." IEEE Trans. Inf. Theor. **11**(1): 61-70.

Shaibu, A. B., B. R. Cho and J. Kovach (2009). "Development of a censored robust design model for time-oriented quality characteristics." Quality and Reliability Engineering International **25**(2): 181-197.
Shumway, R. H. and D. S. Stoffer (1982). "An approach to time series smoothing and forecasting using the EM algorithm." Journal of Time Series Analysis **3**(4): 253-264.
Snyder, R. D. (1985). "Recursive Estimation of Dynamic Linear Models." Journal of the Royal Statistical Society. Series B (Methodological) **47**(2): 272-276.
Spiess, A. N. and N. Neumeyer (2010). "An evaluation of R(2 )as an inadequate measure for nonlinear models in pharmacological and biochemical research: a Monte Carlo approach." BMC Pharmacology **10**: 1-11.
Taylor, S. J. (1987). "Forecasting the volatility of currency exchange rates." International Journal of Forecasting **3**(1): 159-170.
Volterra, V. (1930). Theory of functionals and of integral and integro-differential equations, Blackie & Son Limited.
Wiener, N. (1958). Nonlinear problems in random theory. [Cambridge], Technology Press of Massachusetts Institute of Technology.
Willemain, T. R., C. N. Smart, J. H. Shockor and P. A. DeSautels (1994). "Forecasting intermittent demand in manufacturing: a comparative evaluation of Croston's method." International Journal of Forecasting **10**(4): 529-538.
Wu, B., X. Zhang and R. G. Jeffrey (2014). "A model for downhole fluid and rock temperature prediction during circulation." Geothermics **50**: 202-212.
Yule, G. U. (1927). "On a Method of Investigating Periodicities in Disturbed Series, with Special Reference to Wolfer's Sunspot Numbers." Philosophical Transactions of the Royal Society of London. Series A, Containing Papers of a Mathematical or Physical Character **226**: 267-298.

**Appendix**

**The coefficient of determination ($R^2$):**

The overall accuracy of a regression is measured by the coefficient of determination, $R^2$ which is defined as

$$R^2 = 1 - \frac{SS_{res}}{SS_{tot}} \qquad \ldots\ldots\ldots\ldots\ldots\ldots\ldots\ldots\ldots\ldots (A1)$$

Where,

$SS_{res} = \sum_{i=1}^{n}(Y_{obs,i} - Y_{model,i})^2$ , the residual sum of squares

$SS_{tot} = \sum_{i=1}^{n}(\bar{Y}_{obs} - Y_{model,i})^2$ , the total sum of squares

$\bar{Y}_{obs} = \frac{1}{n}\sum_{i=1}^{n} Y_{obs,i}$ , the mean of observed values

The values of $R^2$ vary from 0 to 1. The $R^2$ value closed to one are indication of better fit of the model curve with observed data.

**Normalized Root Mean Square Error (NRMSE):**

The error in the fitted model is calculated by the difference between the actual or measured value and the predicted value by the model as given in **Equation A.2**

$$Error, e_i = Y_{obs,i} - Y_{model,i} \qquad \ldots\ldots\ldots\ldots\ldots\ldots (A.2)$$

The error calculated from equation A.2 is for a single point. The total error for all points can be measured using mean square error in **Equation A.3**

$$MSE = \frac{\sum_{i=1}^{n} e_i^2}{n} = \frac{\sum_{i=1}^{n}(Y_{obs,i} - Y_{model,i})^2}{n} \quad \ldots\ldots\ldots\ldots\ldots (A.3)$$

The Root Mean Square Error (RMSE) (also known as the root mean square deviation, RMSD), is used to measure the cumulative error for the entire curve.

Often, square root of the MSE or the square root of the mean squared error (RMSE) is used to measure the fitness:

$$RMSE = \sqrt{MSE} \quad \ldots\ldots\ldots\ldots\ldots (A.4)$$

Where $Y_{obs}$ is observed values and $Y_{model}$ is modeled values.

It is not always fair to analyze the error in terms of absolute values because different schemes may have different absolute values and their ranges. Non-dimensional form of the RMSE is used instead by normalizing RMSE with the range of the observed data to obtain Normalized Root Mean Square Error (NRMSE) as given in **Equation A.5**

$$NRMSE = \frac{RMSE}{Y_{obs,max} - Y_{obs,min}} \quad \ldots\ldots\ldots\ldots\ldots (A.5)$$

Where,

$Y_{obs,max}$ is the maximum value of observed data.

$Y_{obs,min}$ is the minimum value of observed data.